\documentstyle[aps,prl,floats,epsf]{revtex}

\begin{document}
\draft
\twocolumn[\hsize\textwidth\columnwidth\hsize\csname@twocolumnfalse%
\endcsname
\title{Two-Dimensional Quantum Ferromagnets}

\author{Carsten Timm$^a$, Patrik Henelius$^a$, Anders W. Sandvik$^b$,
S. M. Girvin$^a$}
\address{$^a$ Department of Physics, Indiana University, Bloomington,
Indiana 47405\\
$^b$ Department of Physics, University of Illinois, 1110 West Green
Street, Urbana, Illinois 61801}
\date{October 22, 1997}
\maketitle
\begin{abstract}
We present $1/N$ Schwinger boson and quantum
Monte Carlo calculations of the magnetization and NMR relaxation
rate  for the two-dimensional ferromagnetic Heisenberg model
representing a quantum Hall system at filling factor $\nu=1$.
Comparing the analytic and numerical calculations, we find that
the SU($N$) version of Schwinger boson theory gives accurate
results for the magnetization at low temperatures,
whereas the O($N$) model works well at higher temperatures.

\end{abstract}
]

Two-dimensional  quantum magnets have received particular attention
in recent years because of advances in materials synthesis associated
with high-temperature superconductivity, thin films and surfaces, and
semiconductor quantum wells.  
It has recently come to be appreciated that
two-dimensional electron gases in quantum wells subjected to strong
magnetic fields in the quantum Hall regime are novel itinerant
ferromagnets.  The strong external magnetic field quenches the orbital
kinetic energy but couples only very weakly to the spin degrees of
freedom allowing low energy spin fluctuations to survive.

Two-dimensional ferromagnets exhibit novel topological defects referred 
to as skyrmions  by analogy with the corresponding objects in the Skyrme
model in nuclear physics.  What is unique about quantum Hall 
ferromagnets \cite{sondhi,cote}
is that these defects carry fermion charge and hence their
ground state density can be controlled by moving the filling factor
away from $\nu=1$.
The combination of low energy spin fluctuations and these topological
defects dramatically alters the NMR spectrum \cite{Barrett,cote}
and the specific heat \cite{bayot}.

With the advent of NMR and magnetoabsorption
measurements in quantum Hall systems, it is now
possible to measure the temperature dependence of the electron
magnetization. Recent theoretical work \cite{RS} has evaluated the
magnetization of this quantum critical system at $\nu=1$ using  
SU($N$) and O($N$) formulations of mean field theory 
(with $N=\infty$). 
In this paper we present analytic results for the $1/N$ corrections
to the magnetization and
compare them with extensive quantum Monte Carlo simulations.
We also present numerical results for the NMR relaxation rate $1/T_1$.
In addition to their relevance
to 2D quantum ferromagnets in general and QHE magnets in particular,
these results provide new information on the level of accuracy of
large $N$ expansion methods and show a highly
non-trivial difference in the behavior of
the SU($N$) and O($N$) models.

Low energy spin fluctuations in quantum Hall ferromagnets at $\nu=1$
are expected to be well described by the Heisenberg model \cite{RS}.
The large $N$ approach to this model is a systematic
expansion around a mean field theory for $N=\infty$ and has
the advantage of being equally valid at all temperatures $T$.
Furthermore, even at the mean field level
this approach correctly captures the fact that arbitrarily small thermal
fluctuations destroy the long range order in two dimensions.
An alternative microscopic approach which includes spin-wave
corrections to the electronic self-energy
has also recently been developed \cite{KM}.
Trumper {\it et al}.\ \cite{TMGC} employ an SU(2) Schwinger boson
theory to describe Gaussian fluctuations in a frustrated 2D
antiferromagnet.

We start from the Heisenberg Hamiltonian
\begin{equation}
H = -J \sum_{\langle ij\rangle} {\bf S}(i)\cdot{\bf S}(j)
  - B \sum_i S^z(i)
\label{sec2H0}
\end{equation}
on a square lattice, where ${\bf S}\cdot{\bf S}=S(S+1)$ at each site.
By introducing two Schwinger bosons at each site,
$S^+ = a^\dagger b$, $S^- = b^\dagger a$, and
$S^z = (a^\dagger a-b^\dagger b)/2$, subject to the constraint
$a^\dagger a + b^\dagger b = 2S$, one obtains an equivalent
boson Hamiltonian. The SU(2) spin algebra of this model is
generalized to SU($N$) \cite{AA}.
After going over to the continuum, the SU($N$) Hamiltonian is
written in terms of $N$ bosons $b_\alpha$ per site,
\begin{eqnarray}
H & = & \int d^2r\, \bigg[ JS\,(\partial_j b_\alpha^\dagger)
    (\partial_j b_\alpha)
  - \frac{J}{N}\, b_\alpha^\dagger(\partial_j b_\beta^\dagger)
    b_\beta(\partial_j b_\alpha)  \nonumber \\
& & {}- \frac{B}{2a^2} h_\beta^\alpha\,b_\beta^\dagger b_\alpha
  \bigg] ,
\label{sec2H1}
\end{eqnarray}
where summation over repeated indices is implied and
$a$ is the lattice constant. The bosons are subject to the
constraint $\sum_\alpha b_\alpha^\dagger b_\alpha = NS$,
which picks out the allowed spin states for total spin $S$.
The matrix $h_\beta^\alpha=\delta_{\alpha\beta} (-1)^{\alpha+1}$
describes the coupling of the bosons to the magnetic field.

The partition function can be written as a coherent state
functional integral over complex fields $b_\alpha$. 
Decoupling the quartic term by means
of a Hubbard-Stratonovich transformation one finds
\begin{displaymath}
Z = \int D^2b_\alpha\,D\lambda\,DQ_j\,
  \exp\!\bigg( -\frac1{\hbar} \int_0^{\hbar\beta}\!\! d\tau
  \int d^2r\: {\cal L}[b;\lambda,{\bf Q}] \bigg)
\end{displaymath}
with the Lagrangian density
\begin{eqnarray}
{\cal L} & = & \frac{\hbar}{a^2}\,b_\alpha^\ast \partial_0 b_\alpha
  + JS\,(\partial_j b_\alpha^\ast)(\partial_j b_\alpha)
  + NJ Q_jQ_j  \nonumber \\
& & {}+ iJ Q_j\,b_\alpha^\ast(\partial_j b_\alpha)
  - iJ Q_j\,(\partial b_\alpha^\ast)b_\alpha
  - \frac{B}{2a^2} h_\beta^\alpha\,b_\beta^\ast b_\alpha  \nonumber
\\
& & {}+ \lambda\,b_\alpha^\ast b_\alpha - NS \lambda ,
\label{sec2L1}
\end{eqnarray}
where $\lambda$ is an imaginary
Lagrange multiplier field which implements
the constraint, and ${\bf Q}$ is the real-valued
Hubbard-Stratonovich field which acts as a gauge field.

The local equivalence of the groups SU(2) and O(3) allows one to
also
write down an O(3) boson model equivalent to Eq.~(\ref{sec2H0}), 
which can be generalized to O($N$) \cite{RS}. The continuum O($N$)
Hamiltonian reads
\begin{eqnarray}
H & = & \int d^2r\, \bigg[ JS\,(\partial_j b_\alpha^\dagger)
    (\partial_j b_\alpha)
  - \frac{3J}{N}\,b_\alpha^\dagger(\partial_j b_\beta^\dagger)
    b_\beta(\partial_j b_\alpha)  \nonumber \\
& & {}- \frac{B}{a^2} h_\beta^\alpha\,b_\beta^\dagger b_\alpha \bigg]
\end{eqnarray}
with the constraints $\sum_\alpha b_\alpha^\dagger b_\alpha = NS/3$
and
$\sum_\alpha b_\alpha^\dagger b_\alpha^\dagger = 0$. The matrix $h$
contains $N/3$ copies of the O(3) generator matrix
$((0,i,0),(-i,0,0),(0,0,0))$
along the diagonal. The second
constraint means that creating two bosons of kind, say, $N$
produces a state which is a linear combination
of states without two $N$ bosons added. This constraint
restricts the Hilbert space by identifying certain states with 
one another.  In the functional integral it
is enforced by a new {\it complex\/} Lagrange multiplier $\mu$,
which enters the Lagrangian as
$\mu^\ast\,b_\alpha b_\alpha/2+\mu\,b_\alpha^\ast b_\alpha^\ast/2$.

To obtain mean field results, the auxiliary fields $\lambda$, ${\bf Q}$,
and, for O($N$), $\mu$ are set to their values at the saddle point.
We denote mean field values by a subscript $0$. After
Fourier transformation, the boson fields can be integrated out.
The values of the auxiliary fields are obtained by solving the
saddle-point equations.
Because of gauge invariance, ${\bf Q}_0$ can be taken to vanish.
For O($N$) we find $\mu_0=0$. The mean field
magnetization $M_0$ is then found from the free energy and
has been given by Read and Sachdev for both models \cite{RS}. 

Another useful quantity is the NMR relaxation rate
\begin{equation}
\frac1{T_1} = \frac{A}{{\cal N}} \sum_{\bf q} S({\bf q}, \omega\to0),
\label{relrate}
\end{equation}
where $A$ is the hyperfine structure factor, which we assume
to be isotropic and momentum independent, ${\cal N}$ is the
number of sites, and $S({\bf q}, \omega)$
is the dynamic structure factor, which is related to the transverse
susceptibility by
$S({\bf q}, \omega) = \text{Im } \chi^{+-}({\bf q},\omega) /
  (1-e^{-\beta\hbar\omega})$.
The mean field rate is plotted in Ref.~\cite{RS},
here, we give analytic results for completeness. For SU($N$),
$1/T_1 = A\,n_B(\Lambda_0+\beta B/2) / (32\pi\beta^2 J^2 S^2)$
and for O($N$), $1/T_1 = A \left[ n_B(\Lambda_0+\beta B) +
n_B(\Lambda_0)
\right] / (16\pi\beta^2 J^2 S^2)$.

We now turn to $1/N$ corrections to the magnetization.
For the SU($N$) case, fluctuations in the
auxiliary fields about their saddle-point values are denoted
by $i\Delta\lambda=\lambda-\lambda_0$ and $\Delta{\bf Q}={\bf Q}$,
respectively. As a short-hand we denote any fluctuation mode by
$r_\ell = \{\Delta\lambda({\bf r},\tau), \Delta Q_j({\bf r},\tau)\}$.
Following the approach outlined in Ref.~\cite{Auerbach},
the partition function is written as a functional integral
over the $r_\ell$,
\begin{equation}
Z = \int Dr_\ell\: e^{-N{\cal S}[r_\ell]}
\label{sec2Z3}
\end{equation}
with the action
${\cal S}={\cal S}_0 + {\cal S}_{\text{dir}} + {\cal
S}_{\text{loop}}$,
where
\begin{eqnarray*}
{\cal S}_0 & = & N^{-1}\,\text{Tr}\, \ln G_0^{-1} , \\
{\cal S}_{\text{dir}} & = & \frac1{N\hbar} \int_0^{\hbar\beta}\!\!
  d\tau \int d^2r \left( NJ {\bf Q}\cdot{\bf Q} - NS\lambda \right) ,
\\
{\cal S}_{\text{loop}} & = & N^{-1}\,\text{Tr}\,
  \ln\left(1+G_0\upsilon_\ell r_\ell\right) .
\end{eqnarray*}
Here, the trace sums over 
momenta, Matsubara frequencies, and boson flavors $\alpha$.
$G_0^\alpha({\bf k},i\omega_n) \equiv (-i\hbar\omega_n + J S k^2a^2
  - B h_\alpha^\alpha/2 + a^2\lambda_0)^{-1}$
is the mean field boson Green function, and the $\upsilon_\ell$ are
vertex factors describing the coupling of the bosons to the
fluctuation $r_\ell$. The action can be written
as a Taylor series in fluctuations $r_\ell$,
${\cal S} = \sum_{n=0}^\infty (n!)^{-1}
  {\cal S}_{\ell_1\ldots\ell_n}^{(n)} r_{\ell_1}\cdots r_{\ell_n}$,
where the $n=1$ term vanishes since we are expanding around
a saddle point. ${\cal S}_0$ conspires with the $r_\ell$
independent part of ${\cal S}_{\text{dir}}$ to form the mean field
free energy $\beta F_0=N{\cal S}^{(0)}$. The other terms in the
series can be found by expanding ${\cal S}_{\text{loop}}$
and ${\cal S}_{\text{dir}}$.
The RPA fluctuation propagator is the inverse of the matrix
${\cal S}^{(2)}$.

\begin{figure}[htb]
\centerline{\epsfxsize 8cm\epsfbox{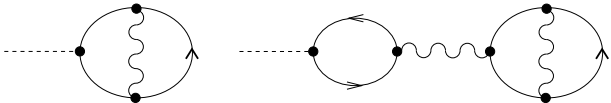}}
\caption{Diagrams of order $1/N$
for $\langle b_\alpha^\dagger b_\alpha\rangle$.
Solid lines denote boson mean field Green functions, wriggly
lines are RPA fluctuation propagators, and
dashed lines are external $j_\alpha$ insertions.}
\label{sec2fig1}
\end{figure}

The magnetization is expressed in terms of boson occupation
numbers, $M=1/N\:\sum_\alpha h_\alpha^\alpha\,
  \langle b_\alpha^\dagger b_\alpha\rangle$,
and the expectation values $\langle b_\alpha^\dagger b_\alpha\rangle$
are obtained by inserting a constant source term
$\sum_\alpha j_\alpha\,b_\alpha^\dagger b_\alpha$ into the
Hamiltonian in Eq.~(\ref{sec2H1}).
We apply this procedure to the partition function of
Eq.~(\ref{sec2Z3}) and expand the result in a Taylor
series in the $r_\ell$. The resulting Gaussian
integrals can be evaluated by contractions over $r_\ell$.
This corresponds to writing down all allowed
diagrams with one external $j_\alpha$ vertex
and any number of internal vertices, connecting all internal
vertices by RPA propagators.
Loops without an external vertex and with one or two internal
ones are forbidden since the former vanish in an expansion
around a saddle point and the latter are included in the RPA
propagator. Fig.~\ref{sec2fig1} shows the diagrams of order $1/N$.

Summation over the frequency of the vertical RPA propagator in both
diagrams must be done carefully,
taking into account normal ordering of operators at equal times.
This procedure makes the frequency sums unambiguous and removes
a spurious divergence. The details will be given elsewhere \cite{CTlong}.
The magnetization can now be calculated numerically.
The momentum integrals have a logarithmic UV divergence, which is
regularized by a lattice cutoff. The $1/N$ contributions from both 
$\Delta\lambda$ and $\Delta Q_j$
decrease the magnetization, as intuitively expected.

For the O($N$) model we additionally have to deal with fluctuations
$\Delta\mu$ in the second Lagrange multiplier field, which couple
only to $b_\alpha^\dagger b_\alpha^\dagger$ and
$b_\alpha b_\alpha$. To order $1/N$ the only contributions come from
the two $1/N$ diagrams with the vertical propagator replaced
by a $\Delta\mu$ RPA propagator and the direction of the boson Green
functions changed accordingly. The result converges for
large momenta and actually increases the magnetization:
Mean field theory, which enforces the
constraint $\sum_\alpha b_\alpha^\dagger b_\alpha^\dagger=0$ only on
average, {\it under\/}estimates the magnetization because it
contains unphysical contributions from spin multiplets of lower
total spin. Analytic results are further discussed below.

In order to test the accuracy of the analytic results, we have carried out 
Quantum Monte Carlo simulations using the 
stochastic series expansion method \cite{sand}, which is ideally suited 
for the present calculation since it does not introduce any systematic errors.
Sufficiently large lattices can be studied so that finite size effects 
are negligible.
The method is based on a Taylor expansion of the density matrix
$e^{-\beta H}$. Writing $H$ is terms of its one- and two-body terms,
$H=\sum_{i=1}^MH_i$, the partition function can be written as \cite{sand}
\begin{equation}
Z = \sum_{\alpha}\sum_{n=0}^{\infty}\sum_{S_n}
  \frac{(-\beta)^n}{n!} \langle\alpha| \prod_{i=1}^n
  H_{l_i} |\alpha\rangle,
\end{equation}
where $S_n$ denotes a sequence of indices $(l_1,l_2,...,l_n)$ with 
$l_i \in {1,...,M}$, and 
$|\alpha \rangle = |S^z_1 ,S^z_2,\ldots, S^z_{\cal N} \rangle$ is an
eigenstate of all the operators $S^z_i$. The relative weight in this
expression can be made positive definite
by adding a suitable constant to $H$. For a system of finite ${\cal N}$
and $\beta$ only sequences of finite length contribute significantly and 
the limit $n \to \infty$ poses no problem (the average power 
$\langle n\rangle$ is given by $|E|\beta$, where $E$ is the total 
internal energy).

We want to emphasize an important feature that makes the 
sampling particularly efficient: The external field is chosen in the 
$\hat x$ direction. This automatically
causes the simulation to become grand-canonical and there are no longer any
problems associated with a restricted winding number. If the transverse field 
is not too weak ($B/J \agt 0.02$), it causes the auto-correlation times of 
all calculated quantities to become very short, even though only purely local 
updates are used. Furthermore, it enables easy access to observables
involving both diagonal and off-diagonal operators. Details of the 
implementation will be presented elsewhere \cite{hene}. 

For a $4\times 4$ system we have compared our QMC data with exact 
diagonalization results, and they agree to within statistical errors.
Relative errors are typically of the order $10^{-4}$ for all system
sizes considered. For all the field strengths presented in this paper
the results for $16\times 16$ and $32\times 32$ sites agree to this 
precision (finite-size effects increase with decreasing $B$), and we here
present magnetization results for the larger size. 

The {\bf q} dependent imaginary time susceptibility is
\begin{displaymath}
\chi^{+-}({\bf q},\tau)={1\over {\cal N}}
\sum_{{\bf r}}\langle S^z({\bf r},\tau)S^z(0,0)\rangle
e^{i{\bf q\cdot r}}.
\end{displaymath}
Once it has been calculated, the dynamic structure factor can 
be obtained by inverting the relation
\begin{displaymath}
\chi^{+-}({\bf q},\tau)={1\over \pi}\int d\omega \, 
{S({\bf q},\omega) }
e^{-\tau\omega}
\end{displaymath}
using the maximum-entropy technique \cite{gube}.
The $B$ field introduces an additional complication to the
analytic continuation since it introduces a delta function
peak in the spectral weight at $\omega =B$ arising from
the ${\bf q} =0$ response. The maximum entropy method cannot
resolve separate delta functions in the spectrum and since we are studying 
relatively small fields this potentially causes problems 
for the calculation of $1/T_1$. We therefore found it useful to 
separate out the ${\bf q}=0$ sector before continuing
the average $\sum_{\bf q} \chi^{+-}({\bf q},\tau)$. The results are, 
however, still sensitive to the somewhat sharp peaks in the 
spectral function near $\omega =B$ from momenta near $q=0$; a problem which
becomes worse with increasing system size. We therefore only use $16\times16$
sites for the $1/T_1$ calculation, which should be sufficient in the 
interesting temperature regime. We estimate the statistical errors using 
the bootstrap technique. As with any results obtained by the maximum entropy 
method, the error estimates have to be viewed with some caution.

\begin{figure}[htb]
\centerline{\epsfxsize 8.5cm\epsfbox{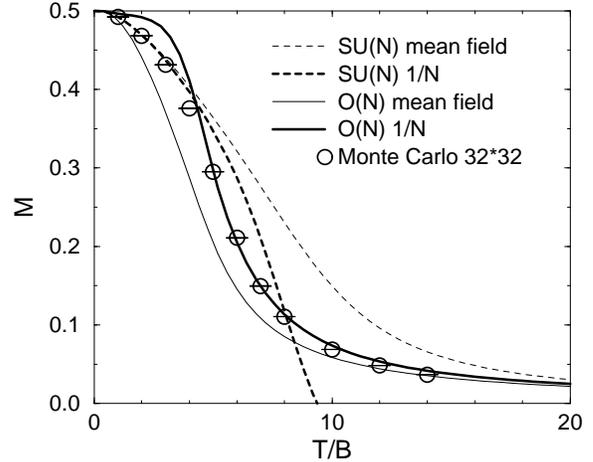}}
\caption{Magnetization of a 2D quantum ferromagnet with
magnetic field $B=0.1\,J$ and spin $S=1/2$.}
\label{sec4fig1}
\end{figure}

Fig.~\ref{sec4fig1} shows the magnetization for spin $S=1/2$ and
magnetic field $B=0.1\,J$. The results are typical for all 
fields considered ($0.02\le B/J \le 0.32$).
The SU($N$) results agree with Monte Carlo data only at the lowest
temperatures. In fact, at low
temperatures the SU($N$) mean field magnetization is known to agree
with the non-interacting magnon approximation for
the Heisenberg model up to exponentially
small corrections. This is not the case for the O($N$) model.
There is a distinct crossover to moderate and high
temperatures, where the O($N$) $1/N$ results agree quite well with
the data, whereas the SU($N$) $1/N$ correction
eventually becomes much too large.
The main differences between the systems studied above
are that (i) the analytic
calculations use a continuum approximation,
and (ii) the leading terms in the
$1/N$ expansion do not include skyrmion effects,
whereas the Monte Carlo simulations in principle do.
Both effects should lead to discrepancies at
temperatures of order $J$ and higher.
We will further pursue the question of why the O($N$) model works
better than the SU($N$) model at most temperatures in a
subsequent paper \cite{CTlong}.

\begin{figure}[htb]
\centerline{\epsfxsize 8.5cm\epsfbox{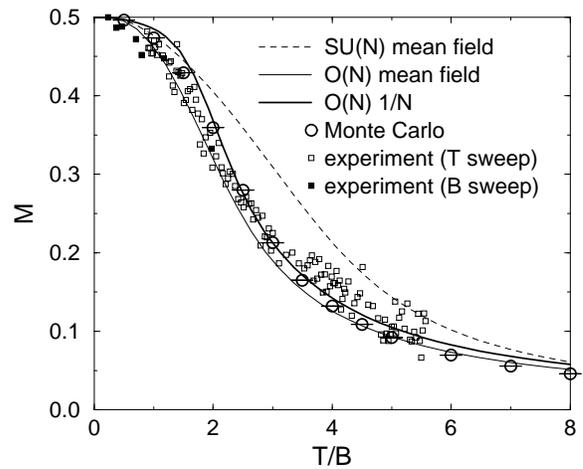}}
\caption{Magnetization for $B=0.32\,J$ compared with experiments
from Ref.~\protect\cite{Manfra}.}
\label{sec4fig2}
\end{figure}

Experimental data by Barrett {\it et al}.~\cite{Barrett}
show a more rapid drop off at higher temperatures, probably
due to disorder affecting the normalization of the data.
In Fig.~\ref{sec4fig2} we compare our results with
recent magnetoabsorption measurements by Manfra
{\it et al}.~\cite{Manfra}. The experiments
agree well with our O($N$) $1/N$ and Monte Carlo results
except at low temperatures. The calculations have been
done for $B=0.32\,J$, which is the experimental field
corrected for finite width of the quantum well \cite{Manfra}.
The approximation used by Kasner and MacDonald \cite{KM}
leads to a much higher magnetization than obtained here.

\begin{figure}[htb]
\centerline{\epsfxsize 8.5cm\epsfbox{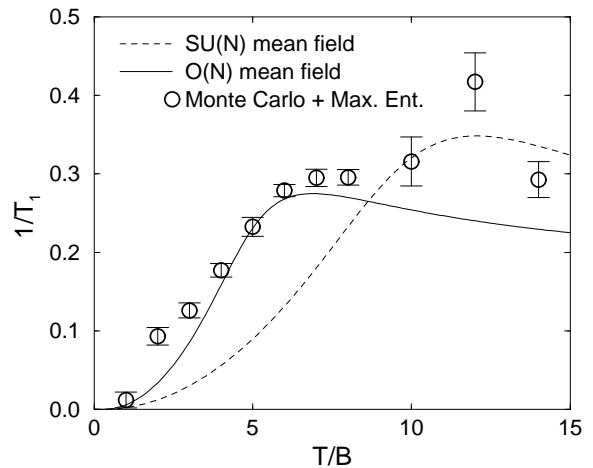}}
\caption{NMR relaxation rate $1/T_1$ for $B/J=0.1$.}
\label{relax}
\end{figure}

In summary, we find from our comparison of Monte Carlo and analytic
approaches that the SU($N$) model correctly captures the
low-temperature magnetization and that the first $1/N$ correction to
the O($N$) model gives accurate results at higher
temperatures. We also find that the NMR relaxation rate evaluated
by analytic continuation from the Monte Carlo data
agrees rather well with the mean field result, as shown in Fig.~\ref{relax}.
Both the $1/N$ and Monte Carlo calculations were rather involved.
Technical details will be presented elsewhere \cite{CTlong,hene}.


We thank S. Sachdev and A. MacDonald for helpful discussions.
The work at Indiana is supported by NSF DMR-9714055 and NSF CDA-9601632,
and at Illinois by NSF DMR-9520776. C.T. acknowledges support by the Deutsche 
Forschungsgemeinschaft. P.H. acknowledges support by the Ella och Georg 
Ehrnrooths stiftelse.

\end{document}